\documentclass[letterpaper]{article} % DO NOT CHANGE THIS
\usepackage{times}  % DO NOT CHANGE THIS
\usepackage{helvet}  % DO NOT CHANGE THIS
\usepackage{courier}  % DO NOT CHANGE THIS
\usepackage[hyphens]{url}  % DO NOT CHANGE THIS
\usepackage{graphicx} % DO NOT CHANGE THIS
\usepackage{caption} % DO NOT CHANGE THIS AND DO NOT ADD ANY OPTIONS TO IT
\usepackage{algorithm}
\usepackage{algorithmic}
\usepackage{fullpage}

\usepackage{newfloat}
\usepackage{listings}
\DeclareCaptionStyle{ruled}{labelfont=normalfont,labelsep=colon,strut=off} % DO NOT CHANGE THIS
\floatstyle{ruled}
\newfloat{listing}{tb}{lst}{}
\floatname{listing}{Listing}
\usepackage{enumerate}
\usepackage{cite}
\usepackage{amsmath,amssymb,amsfonts}
\usepackage{textcomp}
\usepackage{mathtools}

\usepackage{hyperref}       % hyperlinks
\usepackage{booktabs}       % professional-quality tables
\usepackage{bm}
\usepackage{xspace}
\usepackage{xcolor}
\usepackage{color, colortbl}
\usepackage{multirow}
\usepackage{threeparttable}

\newtheorem{proposition}{Proposition}

\newtheorem{theorem}{Theorem}

\newtheorem{assumption}{Assumption}

\def\defn{\,\coloneqq\,}
\def\argmin{\mathop{\mathrm{arg\,min}}} % Argument of a minimization

\def\lim{\mathop{\mathrm{lim}}} % limit

\newcommand{\norm}[1]{\left\lVert#1\right\rVert}

\def\bbm{{\bm{b}}}

\def\ebm{{\bm{e}}}

\def\sbm{{\bm{s}}}
\def\xbm{{\bm{x}}}

\def\ybm{{\bm{y}}}

\def\thetabm{{\bm{\theta}}}

\def\Abm{{\bm{A}}}
\def\Hbm{{\bm{H}}}
\def\Dbm{{\bm{D}}}
\def\Fbm{{\bm{F}}}
\def\Sbm{{\bm{S}}}

\def\Ibm{{\bm{I}}}

\def\Mbm{{\bm{M}}}
\def\Wbm{{\bm{W}}}

\def\xbmhat{{\widehat{\bm{x}}}}

\def\xbmbar{{\bar{\xbm}}}

\def\Tsf{{\mathsf{T}}}

\def\Hsf{{\mathsf{H}}}

\def\C{\mathbb{C}}
\def\R{\mathbb{R}}
\def\E{\mathbb{E}}

\def\Ncal{{\mathcal{N}}}

\definecolor{maroon}{cmyk}{0,0.87,0.68,0.32}

\begin{document}

\title{Self-Supervised Deep Equilibrium Models for Inverse Problems with Theoretical Guarantees}

\date{}

\author{
    Weijie Gan\textsuperscript{\rm 1}, 
    Chunwei Ying\textsuperscript{\rm 3}, 
    Parna Eshraghi\textsuperscript{\rm 2},
    Tongyao Wang\textsuperscript{\rm 2}, 
    Cihat Eldeniz\textsuperscript{\rm 3}, 
    Yuyang Hu\textsuperscript{\rm 4}, \\
    Jiaming Liu\textsuperscript{\rm 4},
    Yasheng Chen\textsuperscript{\rm 5}, 
    Hongyu An\textsuperscript{\rm 2,3,4,5}, 
    Ulugbek S. Kamilov\textsuperscript{\rm 1,4} \\
    % Weijie Gan\textsuperscript{\rm 1}, 
    % Chunwei Ying\textsuperscript{\rm 2}, 
    % Cihat Eldeniz\textsuperscript{\rm 3}, 
    % Tongyao Wang\textsuperscript{\rm 2}, 
    % Yuyang Hu\textsuperscript{\rm 4}, 
    % Jiaming Liu\textsuperscript{\rm 4}, \\
    % Yasheng Chen\textsuperscript{\rm 5}, 
    % Hongyu An\textsuperscript{\rm 2,3,4,5}, 
    % Ulugbek S. Kamilov\textsuperscript{\rm 1,4}\\
    \small \textsuperscript{\rm 1}Department of Computer Science \& Engineering, Washington University in St. Louis, St. Louis \\
    \small \textsuperscript{\rm 2}Department of Biomedical Engineering, Washington University in St. Louis, St. Louis \\
    \small \textsuperscript{\rm 3}Mallinckrodt Institute of Radiology, Washington University in St. Louis, St. Louis \\
    \small \textsuperscript{\rm 4}Department of Electrical \& System Engineering, Washington University in St. Louis, St. Louis \\
    \small \textsuperscript{\rm 5}Department of Neurology, Washington University in St. Louis, St. Louis \\
    \small \texttt{\{weijie.gan,chunwei.ying,p.eshrag,tongyaow,cihat.eldeniz,h.yuyang,}\\
    \small \texttt{jiaming.liu,yasheng.chen,hongyuan,kamilov\}@wustl.edu}
}

\maketitle

\begin{abstract}
  Deep equilibrium models (DEQ) have emerged as a powerful alternative to deep unfolding (DU) for image reconstruction. DEQ models---\emph{implicit neural networks} with effectively infinite number of layers---were shown to achieve state-of-the-art image reconstruction without the memory complexity associated with DU. While the performance of DEQ has been widely investigated, the existing work has primarily focused on the settings where groundtruth data is available for training.
  We present \emph{self-supervised deep equilibrium model (SelfDEQ)} as the first self-supervised reconstruction framework for training model-based implicit networks from undersampled and noisy MRI measurements. Our theoretical results show that SelfDEQ can compensate for unbalanced sampling across multiple acquisitions and match the performance of fully supervised DEQ. Our numerical results on \emph{in-vivo} MRI data show that SelfDEQ leads to state-of-the-art performance using only undersampled and noisy training data.
\end{abstract}

% \begin{IEEEkeywords}
%   Deep learning, deep equilibrium models, inverse problems, magnetic resonance imaging.
% \end{IEEEkeywords}

\section{Introduction}
We consider an \emph{inverse problem} where one seeks to recover an unknown image $\xbm \in \C^n$ from its undersampled and noisy measurements $\ybm \in \C^m$. Inverse problems are ubiquitous across medical imaging, bio-microscopy, and computational photography. In particular, \emph{compressed sensing magnetic resonance imaging (CS-MRI)} is a well known inverse problem that aims to recover diagnostic quality images from undersampled and noisy  \emph{k}-space measurements~\cite{Lustig.etal2007}.
\emph{Deep learning (DL)} has recently gained popularity in inverse problems due to its state-of-the-art performance~\cite{Ongie.etal2020, Lucas.etal2018}.
Traditional DL methods train \emph{convolutional neural networks (CNNs)} to map acquired measurements to the desired images~\cite{Jin.etal2017, Zhu.etal2018}.
Recent work has shown that \emph{deep unfolding (DU)} can perform better than generic CNNs by accounting for the physics of the imaging system~\cite{Aggarwal.etal2019, Schlemper.etal2018}. 
DU models are often obtained from optimization methods by interpreting a \emph{fixed number} of iterations as layers of a deep architecture and training it end-to-end. Despite the empirical success of DU in some applications, the high memory complexity of training DU models limits its use in large-scale imaging applications (e.g., 3D/4D MRI).

Recently, \emph{neural ODEs}~\cite{Chen.etal2018, Kelly.etal2020} and \emph{deep equilibrium models (DEQ)}~\cite{Bai.etal2019, Fung.etal2021} have emerged as frameworks for training deep models with effectively infinite number of layers without the associated memory cost. The potential of DEQ to address imaging inverse problems was recently shown in~\cite{Gilton.etal2021}. Training a DEQ model for inverse problems is analogous to training an \emph{infinite-depth} DU model with constant memory complexity. However, DEQ is traditionally trained using \emph{supervised learning}, which limits its applicability to problems with no groundtruth training data. While there has been substantial interest in developing \emph{self-supervised learning} methods that use undersampled and noisy measurements for training~\cite{zeng2021review,Akcakaya.etal2021,Tachella.etal2022a}, the potential of self-supervised learning has never been explored in the context of DEQ.
This work bridges this gap by proposing \emph{self-supervised deep equilibrium model (SelfDEQ)} as a framework for training \emph{implicit neural networks} for MRI without groundtruth data. 
Our contributions are as follows:
\begin{itemize}
  \item We introduce SelfDEQ as an image reconstruction framework for CS-MRI based on training a model-based implicit neural network directly on undersampled and noisy measurements. SelfDEQ extends the line of work based on \emph{Noise2Noise (N2N)}~\cite{Lehtinen.etal2018} by introducing a model-based implicit architecture, a specialized loss function that accounts for unbalanced sampling, and a memory-efficient training method using \emph{Jacobian-Free Backpropagation (JFB)}~\cite{Fung.etal2021}.
  
  \item We present new theoretical results showing that for certain measurement operators SelfDEQ computes updates that match those obtained by fully-supervised DEQ. In the context of CS-MRI, our results imply that under a set of explicitly specified assumptions, SelfDEQ can provably match the performance of DEQ trained using the groundtruth MRI images. It is worth highlighting that the theoretical guarantees provided by our analysis leverage the proposed correction for unbalanced sampling.
  
  \item We present new numerical results on experimentally-collected \emph{in-vivo} brain MRI data. Our results show that SelfDEQ can (a) outperform recent self-supervised DU methods; (b) match the performance of fully-supervised DEQ, corroborating our theoretical analysis; and (c) enable highly-accelerated data-collection in parallel MRI.
\end{itemize}  

\section{Background}
\subsection{Imaging Inverse Problems}
We consider inverse problems where the measurements $\ybm$ are specified by a linear system
\begin{equation}
  \label{equ:imaging}
  \ybm = \Mbm\Abm\xbm + \ebm\ ,
\end{equation}
where $\xbm$ is the unknown image, $\ebm\in\C^m$ is \emph{additive white Gaussian noise (AWGN)}, $\Abm \in \mathbb{C}^{n \times n}$ is a measurement matrix, and $\Mbm \in \{0, 1\}^{m \times n}$ is a diagonal sampling matrix. A well-known application of~\eqref{equ:imaging} is CS-MRI~\cite{Lustig.etal2007}, where the measurements correspond to the noisy samples in the Fourier domain (referred to as \emph{k}-space).
% In the special case of parallel MRI~\cite{Pruessmann.etal1999}, the transformation matrix can be formulated as $\Abm_i = \Fbm\Sbm_i$ where $\Fbm$ is Fourier transform operator, and $\Sbm_i$ denotes the sensitivity profiles of the $i$th receiver coil.
% We assume $\Sbm$ is known and normalized to satisfy $\sum_i\Sbm_i^\Hsf\Sbm_i=\Ibm$.
% Noted that, in order to estimate $\Sbm$ in practice, $\Pbm$ is required to have \emph{fixed} fully-sampled region in the low-frequency $k$-space, known as \emph{auto calibration signal (ACS)}~\cite{Uecker.etal2014}.

% undersampled measurements at \emph{k}-space $\{\ybm_j\}^M_j$ are obtained from $M$ receiver coils in parallel. Specially, the forward operator of CS-MRI is formulated as 
% \begin{equation}
%   \label{equ:fwd_cs_pmri}
%   \Abm_j=\Pbm\Fbm\Sbm_j\ ,
% \end{equation}
% where $\Fbm\in\C^{n\times n}$ represents the Fourier transform operator, and $\Sbm_j\in\C^{n\times n}$ denotes the sensitivity profiles of the $i$th receiver coil. We assume $\{\Sbm_j\}_j^M$ is known and normalized to satisfy $\sum_j^M\Sbm_j^\Hsf\Sbm_j=\Ibm$. 
% In \eqref{equ:fwd_cs_pmri}, $\Pbm\in\C^{n\times n}$ denotes a diagonal sampling matrix. We consider $\Pbm$ derived from \emph{cartesian} sampling operators that have randomly undersampling and fully sampling trajectories at high- and low-frequency k-space, respectively. The fully-sampling \emph{k}-space data is also known as \emph{auto calibration signal (ACS)}~\cite{Uecker.etal2014}. Noted that ACS is required for estimating $\Sbm$ in parallel MRI. 

Inverse problems are generally ill-posed. Traditional methods recover $\xbm$ by solving a regularized optimization
\begin{equation}
  \xbmhat = \argmin_\xbm f(\xbm) \quad\text{with}\quad f(\xbm) = g(\xbm) + h(\xbm) ,
\end{equation}
where $g$ is the data-fidelity term that quantifies the discrepancy between the measurements and the solution, and $h$ is a regularizer that imposes prior knowledge on the unknown image. Well-known examples in the context of imaging inverse problems are the \emph{least-squares} and \emph{total variation (TV)}
\begin{equation}
  \label{equ:tv}
  g(\xbm)= (1/2)\norm{\ybm-\Mbm\Abm\xbm}_2^2\ \text{and}\ h(x)=\tau\norm{\Dbm\xbm}_1\ ,
\end{equation}
where $\Dbm$ is an image gradient and $\tau > 0$ is the regularization parameter.

\subsection{Deep Learning}
The focus in the area has recently moved to DL (see recent reviews in~\cite{Ongie.etal2020, Lucas.etal2018}). A widely-used DL approach is to train a CNN to learn a mapping from the measurements to the corresponding groundtruth images~\cite{Jin.etal2017, Zhu.etal2018}. There is also a growing interest in \emph{deep model-based architectures (DMBAs)} that can combine physical measurement models and learned image priors specified using CNNs. Well known examples of DMBAs are \emph{plug-and-play priors (PnP)}~\cite{Venkatakrishnan.etal2013, Kamilov.etal2022}, \emph{Regularized by Denoiser (RED)}~\cite{Romano.etal2017}, and \emph{deep unfolding (DU)}~\cite{Schlemper.etal2018, Hammernik.etal2018, Aggarwal.etal2019}. In particular, DU has gained notoriety due to its ability to achieve the state-of-the-art performance, while providing robustness to changes in data acquisition. DU architectures are typically obtained by unfolding iterations of an image reconstruction algorithm as layers, representing the regularizer within image reconstruction as a CNN, and training the resulting network end-to-end. DU architectures, however, are usually limited to a small number of unfolded iterations due to the high memory complexity of training~\cite{Schlemper.etal2018}.

% We contribute to this area by adopting recent developments of deep equilibrium models~\cite{Fung.etal2021} to enable \emph{infinite} iterations in DU. SelfDEQ~can empirically outperform DU methods with much lower memory demand (see also our experimental results on the \emph{in-vivo} data).

\begin{figure*}
  \centering
  \includegraphics[width=.95\textwidth]{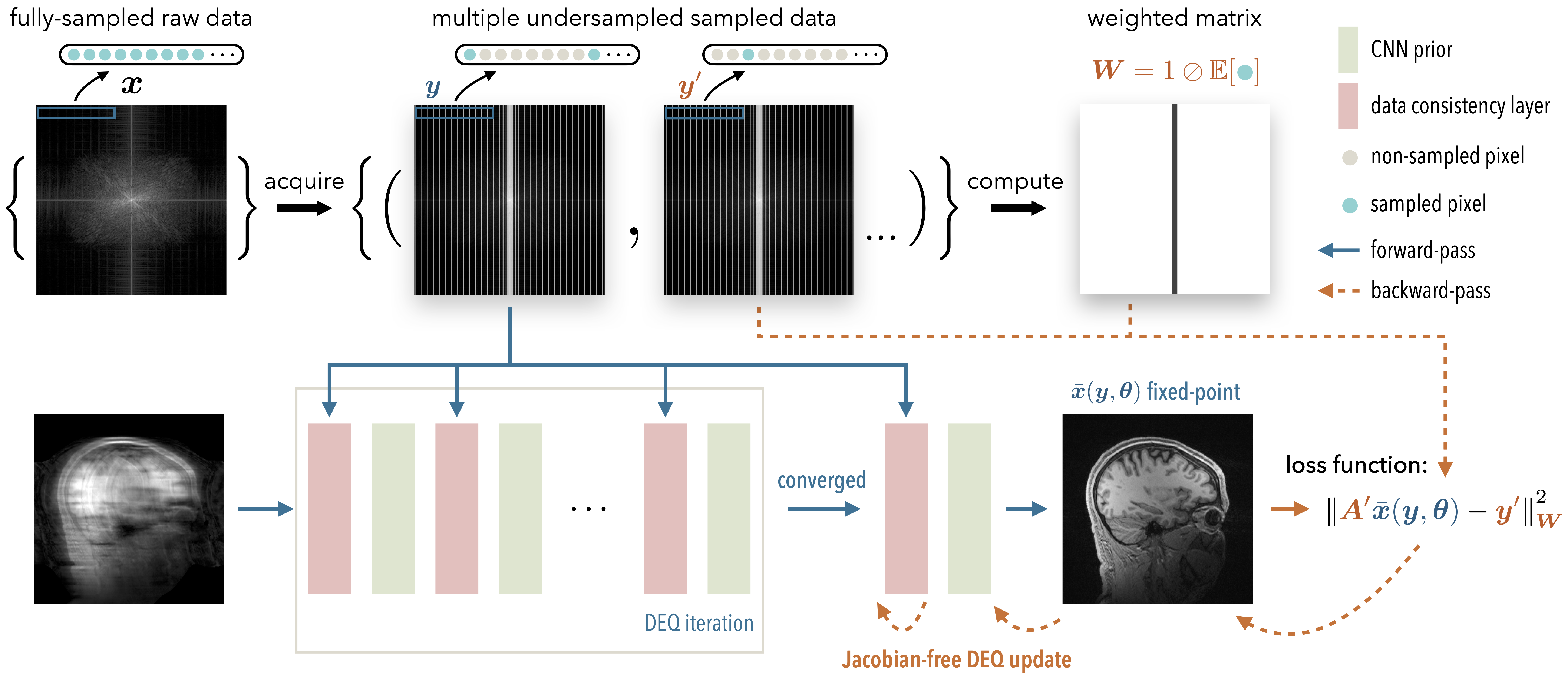}
  \caption{Illustration of SelfDEQ for CS-MRI. The forward pass of SelfDEQ computes a \emph{fixed-point} of an operator consisting of data consistency layer and a CNN prior. The backward pass of SelfDEQ computes a descent direction using the Jacobian-free update that can be used to optimize the training parameters. SelfDEQ is trained using the proposed weighted loss that directly maps pairs of undersampled and noisy measurements of the same object to each other without fully-sampled groundtruth.
%   We impose a weighted matrix in the loss function to compensate over-sampling regions of the physical models. 
%   Our theoretical analysis shows that the proposed DEQ model trained on weighted self-supervised loss function can achieve equivalent performance against the supervised counterpart.
  }
  \label{fig:method}
\end{figure*}

\subsection{Deep Equilibrium Models}
DEQ has emerged as a framework for training recursive networks that have \emph{infinitely} many layers without storing intermediate latent variables~\cite{Bai.etal2019, Fung.etal2021, Pramanik.Jacob2022a, Pramanik.Jacob2022, Gilton.etal2021, Zhao.etal2022, Liu.etal2022}. 
It is implemented by running two consecutive steps in each training iteration, namely the \emph{forward} pass and the \emph{backward} pass. The forward pass computes a fixed point $\xbmbar$ of an operator $\mathsf{T}_\thetabm$ parameterized by weights $\thetabm$
\begin{equation} 
  \label{bk:fixed_point}
  \xbmbar = \mathsf{T}_\thetabm(\xbmbar, \ybm)\ ,
\end{equation}
where $\ybm$ is the measurement vector. The fixed point $\xbmbar$ is often computed by running a fixed-point iteration with an acceleration algorithm (such as Anderson acceleration~\cite{Anderson1965}). It is worth noting that, when $\mathsf{T}_\thetabm$ denotes a step of DMBA, the DEQ forward pass is equivalent to DU with infinitely many unfolded layers. Given a loss function, the backward pass produces gradients with respect to $\thetabm$ by implicitly differentiating through the fixed points without the knowledge of how they are estimated (see Sec.~4.2 in~\cite{Gilton.etal2021} for more details).
% \begin{equation}
%   \label{equ:deq_general_backward}
%   \frac{\partial \ell}{\partial \thetabm} = \frac{\partial \mathsf{T}_\thetabm(\xbmbar, \ybm)}{\partial \thetabm}\ \bar{\bbm}\ \frac{\partial \ell}{\partial \xbmbar} \ \Rightarrow\ \bar{\bbm}={\Big[\Ibm - \frac{\partial \mathsf{T}_\thetabm(\xbmbar, \ybm)}{\partial \xbm}\Big]}^{-\mathsf{T}}
% \end{equation}
% Here, $\Ibm$ is an identity matrix, and the vector product with the inverse-Jacobian $\bar{\bbm}$ can be approximated by solving a fixed-point iteration (see also Sec.4.2 in~\cite{Gilton.etal2021} for more details).
DEQ does not require storing the intermediate variables for computing the gradient, which dramatically reduces the memory complexity of training. 
There have been several applications of DEQ in imaging, including applications to MRI~\cite{Pramanik.Jacob2022a, Pramanik.Jacob2022, Gilton.etal2021}, computed tomography (CT)~\cite{Liu.etal2022} and video snapshot imaging~\cite{Zhao.etal2022}. 
% For example,~\cite{Pramanik.Jacob2022a} and~\cite{Pramanik.Jacob2022} exploit monotone operator~\cite{Winston.Kolter2020} to provide guaranteed convergence and robustness to input perturbations.~\cite{Liu.etal2022} proposes a new strategy for improving the efﬁciency of DEQ through stochastic approximations of the measurement models.
% Note that SelfDEQ~focuses on training strategy and thus is fully compatible with contemporary architectures of DEQ, such as~\cite{Liu.etal2022}.
 
\subsection{Self-Supervised Deep Image Reconstruction}
\label{sec:related-work} 
There is a growing interests in developing DL methods that reduce the dependence on the groundtruth training data (see recent reviews~\cite{zeng2021review, Akcakaya.etal2021,Tachella.etal2022a}). Some well-known strategies include \emph{Noise2Noise (N2N)}~\cite{Lehtinen.etal2018}, \emph{Noise2Void (N2V)}~\cite{Krull.etal2019}, \emph{deep image prior (DIP)}~\cite{Ulyanov.etal2018}, \emph{Compressive Sensing using Generative Models (CSGM)}~\cite{gupta2021cryogan, bora2018ambientgan}, and equivariant imaging~\cite{chen2021equivariant}. 
In particular, N2N is one of the most widely-used self-supervised DL frameworks for image restoration that directly uses noisy observations $\{\xbmhat_{i,j} = \xbm_i+\ebm_{i,j}\}$ of groundtruth images $\{\xbm_i\}$ for training.
The N2N training can be formulated as
\begin{equation}
  \label{equ:n2n}
  \argmin_\thetabm \sum_i\sum_{j\neq j'}\norm{\mathsf{f}_\thetabm(\xbmhat_{i,j}) - \xbmhat_{i,j'}}^2_2\ ,
\end{equation}
where $\mathsf{f}_\thetabm$ denotes the DL model with trainable parameters $\thetabm$. There have been many extensions of N2N to different imaging problems, such as MRI~\cite{Yaman.etal2020, Millard.Chiew2022, Gan.etal2022}, OCT~\cite{Jiang.etal2021}, and CT~\cite{Hendriksen.etal2020}.
In particular, SSDU~\cite{Yaman.etal2020} is a recent state-of-the-art method based on training a DU model without groundtruth by dividing a single k-space MRI acquisition into two subsets that are used as training targets for each other. The work~\cite{Millard.Chiew2022} has provided a theoretical justification for SSDU by extending Noisier2Noise~\cite{Moran.etal2020} to variable-density subsampled MRI data. 

\subsection{Our Contributions}
While DEQ has been shown to achieve the state-of-the-art imaging performance, the existing work has focused on settings where groundtruth data is available for training. Our work addresses this gap by enabling DEQ training on noisy and undersampled sensor measurements, which has not been investigated before. The proposed SelfDEQ framework consists of several synergistic elements: (a) a model-based implicit network that integrates measurement operators and CNN priors; (b) a self-supervised loss that accounts for sampling imbalances; (c) a Jacobian-free backward pass that leads to efficient training.

%The key difference of our work is that it extends N2N and establishes the theoretical analysis in the context of DEQ.
%To to specific, our work enables training DEQ directly from noisy measurements and gains theoretically gain comparable performance against the supervised learning by using a weighted loss function.
% While A2A/DeCoLearn were validated empirically, this study represents a theoretical analysis to justify the proposed self-supervised learning. Moreover, we propose a \emph{weighted} loss function to compensate oversampling regions between different acquisitions.

\section{Method}

\subsection{Weighted Self-Supervised Loss}
\label{sec:loss}

Consider the training set of measurement pairs $\{\ybm_i, \ybm_i'\}_{i = 1}^N$ with each pair $\ybm_i, \ybm_i'$ corresponding to the same object $\xbm_i$
\begin{equation}
  \label{equ:proposed_imging}
  \ybm_i = \Mbm_i\Abm\xbm_i + \ebm_i\ \text{and}\ \ybm_i'=\Mbm_i'\Abm\xbm_i+\ebm_i'\ .
\end{equation}
Here, $N \geq 1$ denotes the number training pairs. 
One can obtain measurement pairs by physically conducting two acquisitions or splitting each acquisition into two subsets.

Existing algorithms based on N2N directly map the measurement pairs to each other during training. However, the measurements in the training dataset often have a significant overlap. For example, each acquisition may share the \emph{auto calibration signal (ACS)} region~\cite{Uecker.etal2014}, thus giving more weight to corresponding regions of the k-space (see SelfDEQ~\emph{(unweigted)} in Fig.~\ref{fig:real_result_aba}).
We introduce a diagonal weighted matrix $\overline{\Wbm}=\mathsf{diag}(\overline{w_0}, \overline{w_1}, ..., \overline{w_{n}})\in\R^{n\times n}$ that accounts for the oversampled regions in the loss function.
We set the diagonal entries of $\overline{\Wbm}$ as follows
\begin{equation}
    \overline{w_k} = \begin{cases}
  \frac{1}{\sqrt{{\E[{\Mbm'}^\Tsf\Mbm']}_{k,k}}}&{\sqrt{\E[{\Mbm'}^\Tsf\Mbm']_{k,k}}} \neq 0 \\
  0&\sqrt{{\E[{\Mbm'}^\Tsf\Mbm']}_{k,k}} = 0
\end{cases}\ ,
\end{equation}
where, in practice, the expectation over random sampling patterns can be replaced with an empirical average over the training set.
We can then define the following self-supervised training loss function
\begin{equation}
  \label{equ:self}
  \ell_{\mathsf{self}}(\thetabm) = \E\norm{\Mbm'\Abm'\xbmbar(\thetabm) - \ybm'}^2_{\Wbm}\ ,
\end{equation}
where $\Wbm=\Mbm'\overline{\Wbm}{(\Mbm'\overline{\Wbm})}^\Tsf\in\R^{m\times m}$ denotes a subsampled variant of $\overline{\Wbm}$ given $\Mbm'$, and
$\xbmbar = \mathsf{T}_\thetabm(\xbmbar, \ybm)$ denotes the fixed-point of $\mathsf{T}_\thetabm$ for the input $\ybm$ and weights $\thetabm$.

% \begin{figure}
%   \centering
%   \includegraphics[width=.48\textwidth]{Fig/Fig_network.pdf}
%   \caption{Architecture of $\mathsf{f}_\thetabm$. We set $C$ to 64 in practice. One can find the detailed implementation in our published code. Best viewed in digital.}
%   \label{fig:network}
% \end{figure}

% This section represents SelfDEQ~.We start by depicting the forward-backward pass of SelfDEQ~and then discuss the associated self-supervised learning strategy, followed by the theoretical analysis. Fig.~\ref{fig:method} illustrates the pipeline of SelfDEQ.

\subsection{Forward and Backward Passes}
The SelfDEQ forward pass is a fixed-point iteration
\begin{equation}
  \label{equ:forward_pass}
  \xbm^k = \mathsf{T}_\thetabm(\xbm^{k-1}, \ybm)\ ,
\end{equation}
where
\begin{equation}
  \begin{aligned}
    \mathsf{T}_\thetabm(\xbm) & = 
     \alpha\mathsf{f}_\thetabm(\sbm) + (1-\alpha) \sbm\ \\
     &\mathsf{with}\ \sbm=\xbm - \gamma\nabla g(\xbm)
  \end{aligned}\
\end{equation}
The vector $\xbm^k$ denotes the image at the $k$th layer of the implicit network, $\gamma$ and $\alpha$ are two hyper-parameters, and $\mathsf{f}_\thetabm$ is the CNN prior with trainable parameters $\thetabm$. The implicit neural network is initialized using the pseudoinverse of the raw measurements  $\xbm^0=\Abm^\dagger\ybm$, which corresponds to the zero-filled solution in CS-MRI.
Given $\ybm$, we run the forward-pass until convergence to a fixed-point $\xbmbar = \mathsf{T}_\thetabm(\xbmbar, \ybm)$. We use U-Net for $\mathsf{f}_\thetabm$~\cite{Ronneberger.etal2015}.
% (see detailed architecture in Fig.~\ref{fig:network}). 
%Especially, spectral normalization is added at outputs of all the convolutional layers in $\mathsf{f}_\thetabm$ to guarantee the fixed-point convergence~\cite{{Ryu.etal2019}}.

Given the self-supervised loss in eq.~\eqref{equ:self}, the SelfDEQ backward pass uses a Jacobian-free backward pass (JFB) to compute the DEQ update direction without computing the inverse-Jacobian. Specifically, the JFB update of $\ell_{\textsf{self}}$ in term of $\thetabm$ is given by
\begin{equation}
  \label{equ:jbf}
  \mathsf{JFB}_{\ell_\mathsf{self}}(\thetabm) = \mathsf{Real}\Big(\big(\nabla_\thetabm\Tsf_\thetabm(\xbmbar)\big)^\Hsf{\Big[\frac{\partial\ell_{\textsf{self}}}{\partial\xbmbar}\Big]}^\Tsf\Big)\ .
\end{equation}
% Disengaging $\bar{\bbm}$ from DEQ gradients can reduce the computational and memory complexity in the backward pass. 
% The JFB gradient is in form equivalent to the gradient of non-DEQ DL models except requiring the fixed-points as the predictions. 
% Therefore, it is relatively easy to implement JFB gradient in practice. 
JFB was theoretically shown to provide valid descent directions for training implicit networks in~\cite{Fung.etal2021}.

\subsection{Theoretical Analysis}

We now present the theoretical analysis of SelfDEQ learning under two explicitly-specified assumptions. 
\begin{assumption}
  \label{AS:1}
  \emph{The training samples correspond to the setting in~\eqref{equ:proposed_imging} with $\xbm\sim p_\xbm$, $\Mbm\sim p_\Mbm$, $\Mbm'\sim p_\Mbm$, $\ebm\sim \Ncal(0, \sigma^2)$ and $\ebm'\sim \Ncal(0, \sigma^2)$ drawn i.i.d. from their respective distributions.}
\end{assumption}
This mild assumption simply states that the sampling matrices, images, and noise are all sampled independently from each other, which is a reasonable assumption.
\begin{assumption}
  \label{AS:2}
   \emph{$\E_{\Mbm}[{\Mbm}^\Tsf\Mbm]$ has a full rank and $\Abm$ is an orthogonal matrix, where the expectation is taken over $p_{\Mbm}$.}
\end{assumption} 
This assumption implies that \emph{union} of all the sampling matrices $\{\Mbm\}$ covers the complete measurement domain. Note that each individual $\Mbm$ can still be undersampled.
\begin{theorem}
\label{THE}
  Under Assumptions~1-2, the JFB update of the weighted self-supervised loss~\eqref{equ:self} is equivalent to its supervised counterpart, namely we have that
\begin{equation}
    \begin{aligned}
        & \mathsf{JFB}_{\ell_\mathsf{self}}(\thetabm) = \mathsf{JFB}_{\ell_\mathsf{sup}}(\thetabm)\ .
    \end{aligned}
\end{equation}
where 
\begin{equation}
\label{equ:supervised}
    \ell_\mathsf{sup}(\thetabm)=\E\left[\frac{1}{2}\norm{\xbmbar(\thetabm) - \xbm}_2^2\right].
\end{equation}
\end{theorem}
The proof is provided in the supplementary material. Theorem~\ref{THE} states that the JFB updates from our weighted self-supervised loss theoretically match those obtained using conventional supervised learning on DEQ. These updates can be easily integrated into any DL optimization algorithm, such as SGD and Adam~\cite{Kingma.Ba2017}.

\begin{figure}
  \centering
  \includegraphics[width=.45\textwidth]{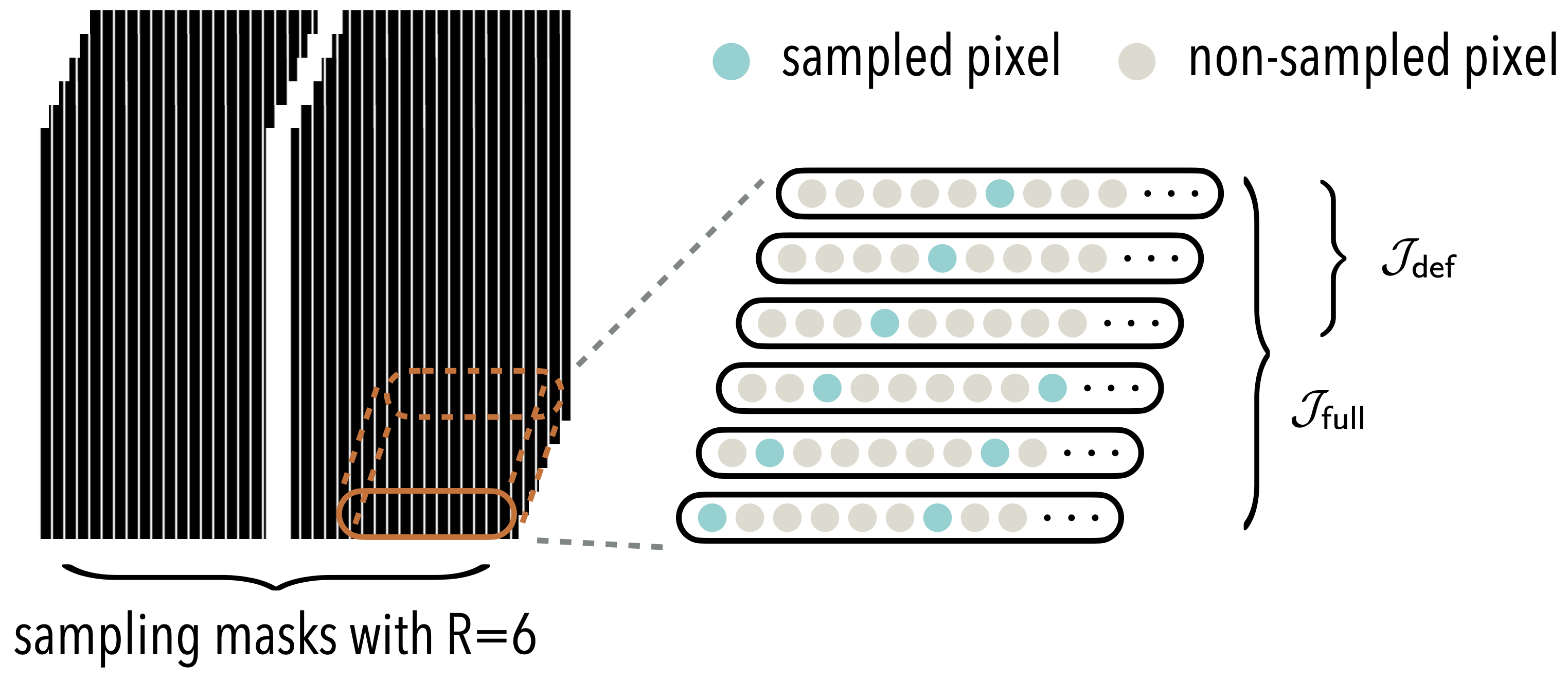}
  \caption{An illustration of sampling masks used for our numerical evalution for the acceleration factor $R = 6$. We consider two settings: full-rank (${\mathcal{J}_\mathsf{full}}$) and rank-deficient ($\mathcal{J}_\mathsf{def}$). In the full-rank setting, the union of sampling masks across training data covers the whole space $\C^n$, thus satisfying Assumption~\ref{AS:2}. In the rank-deficient setting, some of the frequencies are never sampled in the training dataset. Note that each individual sampling mask is always undersampled.
  }
  \label{fig:simulated_mask}
\end{figure}

\section{Numerical Validation}
\label{sec:exp}
We now presents numerical results evaluating SelfDEQ on both simulated and \emph{in-vivo} MRI data. The measurement matrix in parallel MRI can be expressed as $\Abm_i=\Fbm\Sbm_i$ where $\Fbm$ is the Fourier transform operator, and $\Sbm_i$ denotes the sensitivity profiles of the $i$th receiver coil.
We assume that $\Sbm$ is known and normalized to satisfy $\sum_i\Sbm_i^\Hsf\Sbm_i=\Ibm$. Since the Fourier transform operator is orthogonal, the matrix $\Abm$ in MRI naturally satisfies Assumption~\ref{AS:2}. 
Note that, in order to estimate $\Sbm_i$ in practice, $\Mbm$ has a \emph{fixed} ACS in the low-frequency region of $k$-space~\cite{Uecker.etal2014}. 
The random valuables in $p_\Mbm$ in our experiments are the randomly sampled non-ACS indices of the $k$-space.

We ran the forward-pass of SelfDEQ with a maximum number of iterations of 100 and the stopping criterion of the relative norm difference between iterations being less than $10^{-3}$.
We added spectral normalization to all the layers of $\mathsf{f}_\thetabm$ for stability~\cite{Ryu.etal2019}.
We empirically determined the best values of $\alpha$ and $\gamma$ to be $\alpha = 0.5$ and $\gamma=1$.
We used Adam~\cite{Kingma.Ba2017} as the optimizer with the learning rate $10^{-4}$.
We set the mini-batch size to 8 and training epochs to 100 and 300 for real and simulated data, respectively.
We performed all our experiments on a machine equipped with an AMD Ryzen Threadripper 3960X Processor and an NVIDIA GeForce RTX 3090 GPU.  We used two widely used quantitative metrics, \emph{peak signal-to-noise ratio (PSNR)} measured in dB and \emph{structural similarity index (SSIM)}, to evaluate the quality of reconstructed images.
% Our code is available at \href{https://github.com/}{github.com/wustl-cig/SS-DEQ}.

\begin{figure*}[!ht] 
  \centering
  \includegraphics[width=.95\textwidth]{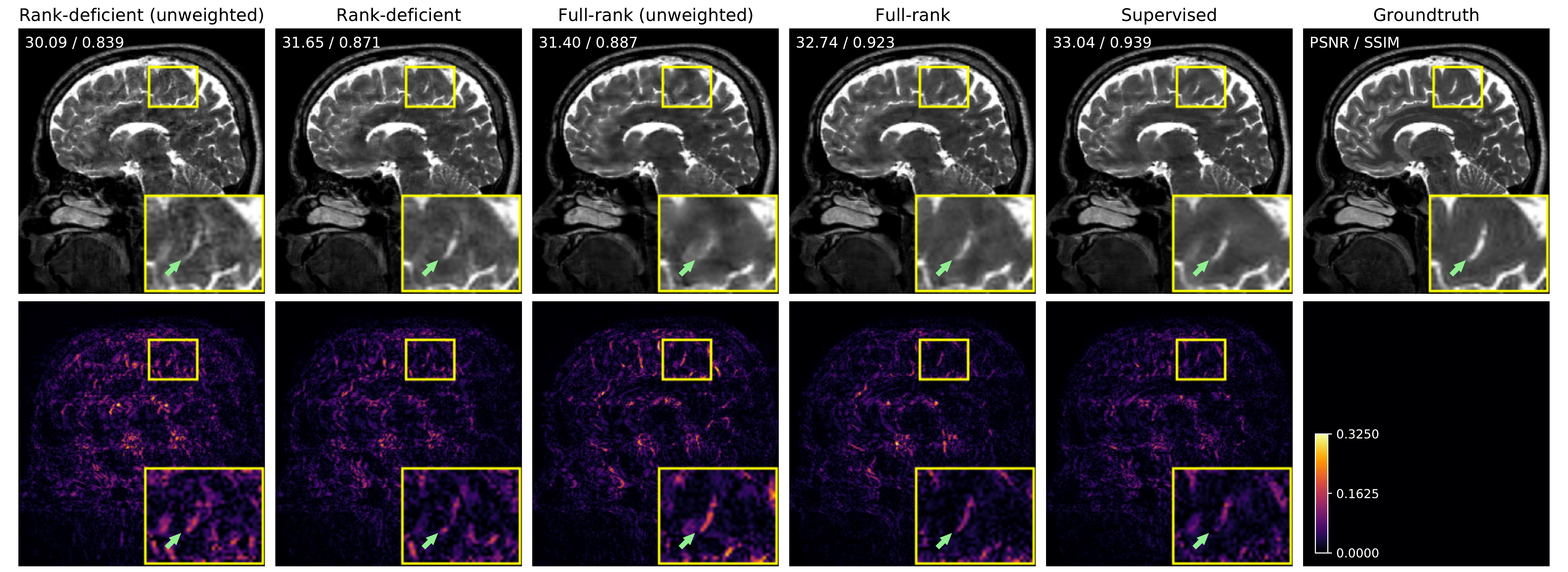}
  \caption{An illustration of reconstructed results obtained from several ablated variants of SelfDEQ~using the acceleration factor $R=6$ on simulated data. We highlight the visually significant differences using green arrows. This figure highlights that in the \emph{Full-rank} setting with the weighting matrix in the loss function, SelfDEQ nearly matches the performance of its supervised counterpart. This figure also shows that the proposed weighting matrix within the self-supervised loss significantly improves the imaging quality, even when Assumption~\ref{AS:2} is not satisfied (compare \emph{Rank-deficient (unweighted)} and \emph{Rank-deficient}).
  }
  \label{fig:simulated_result}
\end{figure*} 

\begin{figure*}[!ht]
  \centering
  \includegraphics[width=.95\textwidth]{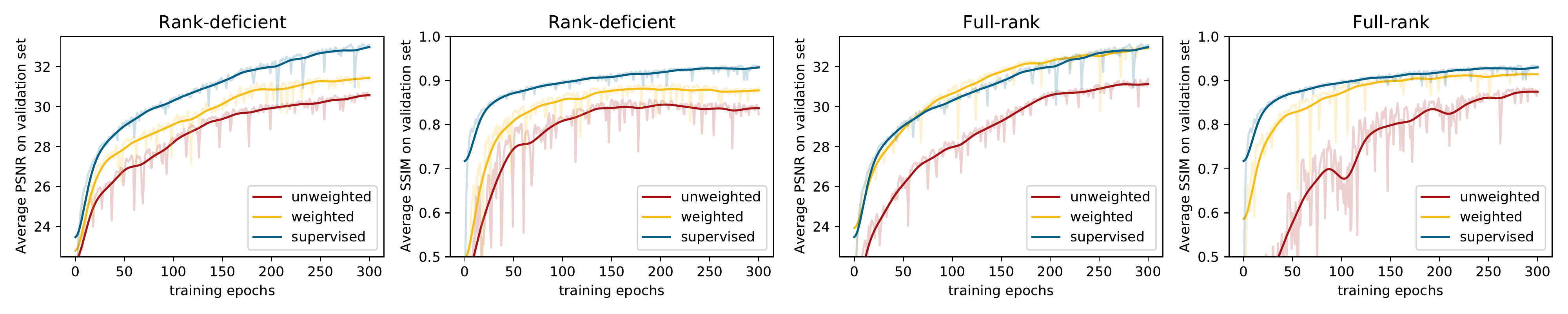}
  \caption{PSNR/SSIM plotted against the training epochs using the simulated validation dataset. This figure highlights that performance of SelfDEQ in the \emph{Full-rank} setting with a weighting matrix closely tracks that of the \emph{supervised} DEQ. The figure also shows the benefit of using the weighting matrix for self-supervised training in both \emph{Rank-deficient} and \emph{Full-rank} scenarios.
  }
  \label{fig:simulated_plot}
\end{figure*}

\subsection{Ablation Study on Simulated Data}
\subsubsection{Dataset} We simulated multi-coil undersampled measurements from an open-access T2-weighted human brain MRI data\footnote{This dataset is available at \url{https://drive.google.com/file/d/1qp-l9kJbRfQU1W5wCjOQZi7I3T6jwA37/view?usp=sharing}}, which was collected by~\cite{Aggarwal.etal2019}.
This MRI dataset has 360 and 160 slices of fully-sampled \emph{k}-space measurements for training and testing, respectively. 
We extracted 60 slices from the training set for validation.
The image domain matrix size for each slice is $256\times 232$. The number of receiver coils is 12. The coil sensitivity maps for each slice are also provided, which are pre-computed using ESPIRiT algorithm~\cite{Uecker.etal2014}.
These fully-sampled data correspond to the groundtruth in the imaging system (\emph{i.e.,} $\xbm_i$ in \eqref{equ:proposed_imging}).
We simulated a Cartesian sampling pattern that subsamples and fully samples along $k_y$ and $k_x$ dimensions, respectively. The simulated sampling mask in the $ky$ dimension has fixed ACS lines and equispaced non-ACS lines. Let $R$ be the acceleration factor. We set the size of ACS lines to $92 / R$. We conducted experiments on three acceleration factors $R=4, 6,\text{ and }8$, corresponding to $31\%$, $21\%$ and $16\%$ sampling rate, respectively. 
As illustrated in Fig.~\ref{fig:simulated_mask}, given a acceleration factor $R$, one can simulate $R$ different sampling masks. Let $\mathcal{J}$ denote a \emph{subset} of those simulated masks. When simulating the measurements, we sampled sampling masks uniformly at random from $\mathcal{J}$ (\emph{i.e.}, $\Mbm_i$ and $\Mbm'_i$ in \eqref{equ:proposed_imging}).
We set the standard deviation of the AWGN (\emph{i.e.}, $\ebm_i$ and $\ebm'_i$ in \eqref{equ:proposed_imging}) to $0.01$.

\subsubsection{Results}
We consider two different sampling settings: full-rank ($\mathcal{J}_\mathsf{full}$) and rank-deficient ($\mathcal{J}_\mathsf{def}$). In the full-rank setting the sampling masks across training data cover all possible frequencies, while in the rank-deficient setting the union of all sampling masks only covers \emph{half} of the k-space. Fig.~\ref{fig:simulated_mask} visually illustrates both settings $\mathcal{J}_\mathsf{full}$ and $\mathcal{J}_\mathsf{def}$ for the acceleration factor $R = 6$. Note that the sampling masks selected from $\mathcal{J}_\mathsf{full}$ naturally satisfy Assumption~\ref{AS:2}, since $\E_{\Mbm\in\mathcal{J}_\mathsf{full}}[\Mbm^\Tsf\Mbm]$ has full rank.
On the other hand, $\E_{\Mbm\in\mathcal{J}_\mathsf{def}}[\Mbm^\Tsf\Mbm]$ does not satisfy Assumption~\ref{AS:2}.
% One the other hand, while the sampling masks selected from $\mathcal{J}_\mathsf{def}$ does not satisfy Assumption~\ref{AS:2}, they are more aligned to practical acquisition scenarios as well as the experimental setup of our real data (see also the next section). 
Under these two sampling settings, we ran the following experiments using ablated variants of SelfDEQ: (a) \emph{Rank-deficient} trains SelfDEQ on  $\mathcal{J}_\mathsf{def}$; (b) \emph{Rank-deficient (unweighted)} is similar to \emph{Rank-deficient}, but uses the self-supervised loss without $\Wbm$; (c) \emph{Full-rank} trains SelfDEQ~on $\mathcal{J}_\mathsf{full}$; (d) \emph{Full-rank (unweighted)} is similar to \emph{Full-rank}, but uses the self-supervised loss without $\Wbm$; (e) \emph{Supervised} is similar to \emph{Full-rank} but uses the supervised loss in \eqref{equ:supervised}, corresponding to the oracle DEQ performance.

% \begin{enumerate}[\ (a)]
%   \item \emph{Rank-deficient}: trains SelfDEQ~on the setup of $\mathcal{J}_\mathsf{def}$;
%   \item \emph{Rank-deficient (unweighted)}: this method is similar to \emph{Rank-deficient}, but uses the self-supervised loss function without the weighted matrix $\Wbm$;
%   \item \emph{Full-rank}: trains SelfDEQ~on the setup of $\mathcal{J}_\mathsf{full}$;
%   \item \emph{Full-rank (unweighted)}: is similar to \emph{Full-rank}, but uses the self-supervised loss function without the weighted matrix $\Wbm$;
%   \item \emph{Supervised}: is similar to \emph{Full-rank} but use the supervised loss function shown in \eqref{equ:supervised}, corresponding to the oracle performance that the proposed DEQ can achieve.
% \end{enumerate} 

\begin{figure*}[!ht]
  \centering 
  \includegraphics[width=.95\textwidth]{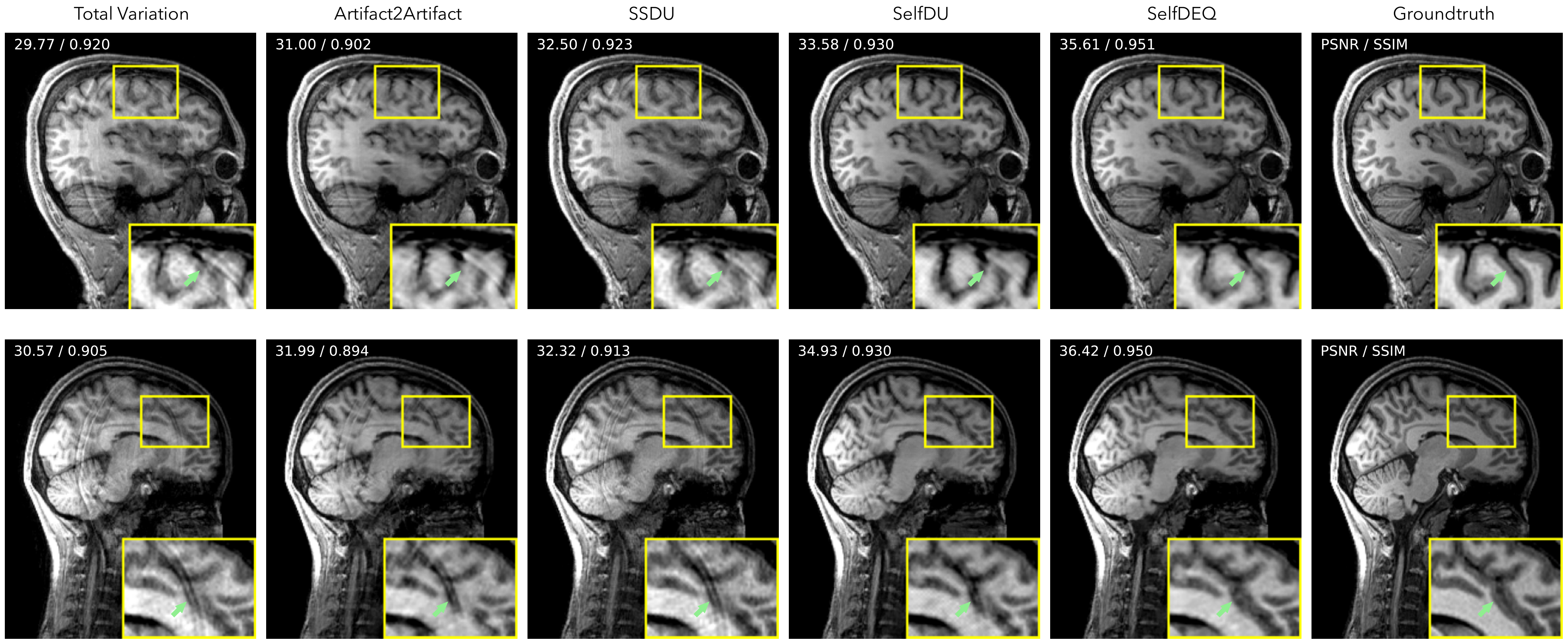}
  \caption{Reconstruction results on in-vivo data comparing several reconstruction algorithms on two different slices under the acceleration factor $R=6$. Note that all the DL methods in the figure are based on self-supervised learning. We highlight visually significant differences with green arrows. Note the regions where SelfDEQ~reconstructs images with fine details, but other methods result in ghosting artifacts. This figure highlights that SelfDEQ can achieve superior performance compared to recent self-supervised DU methods on experimentally-acquired parallel MRI data.
  }
  \label{fig:real_result}
\end{figure*} 

\begin{table}[!ht]
  \centering
  \small
  \renewcommand\arraystretch{1.1}
  \setlength{\tabcolsep}{3.7pt}
  \begin{threeparttable}
  \begin{tabular}{ccccccc}
  \toprule
  {Metrics}                   & \multicolumn{3}{c}{PSNR (dB)}       & \multicolumn{3}{c}{SSIM}    \\
  {Acceleration rate}             & $\times$8   & $\times$6   & $\times$4    & $\times$8   & $\times$6   & $\times$4    \\
  \cmidrule(rl){2-4}\cmidrule(rl){5-7}
  \emph{Zero-Filled}               & 22.27 & 23.40 & 26.59 & 0.771 & 0.798 & 0.861 \\
  \rowcolor{gray!20}\emph{Rank-deficient}$^\mathsf{w/o}$ & 29.58 & 31.86 & 38.54 & 0.843 & 0.856 & 0.905 \\
  \emph{Rank-deficient}  & 30.31 & 33.45 & 39.91 & 0.880 & 0.894 & 0.936 \\
  \rowcolor{gray!20}\emph{Full-rank}$^\mathsf{w/o}$      & 29.95 & 32.73 & 38.61 & 0.861 & 0.892 & 0.922 \\
  \emph{Full-rank}       & {\bf 31.28} & {\bf 34.14} & {\bf 41.56} & {\bf 0.903} & {\bf 0.931} & {\bf 0.951} \\
  \midrule
  \emph{Supervised}                & 32.35 & 34.68 & 42.02 & 0.925 & 0.955 & 0.981 \\
  \bottomrule 
  \end{tabular}
  \begin{tablenotes}
  \item \emph{Rank-deficient}$^\mathsf{w/o}$: \emph{Rank-deficient (unweighted)};
  \item \emph{Full-rank}$^\mathsf{w/o}$: \emph{Full-rank (unweighted)}.
  \end{tablenotes}
\end{threeparttable}
\caption{Average PSNR and SSIM values from the ablation study on simulated data. \emph{Full-rank} satisfies assumptions used for the theoretical analysis and is trained using weighted self-supervised loss. This table validates our theoretical analysis and highlights the importance of the weighting for the self-supervised loss function in both rank-deficient and full-rank settings.}
\label{tb:simulated_result}
\end{table}

\subsubsection{Discussion}
Fig.~\ref{fig:simulated_result} illustrates the reconstruction results of all the ablated methods on $R = 6$. Fig.~\ref{fig:simulated_result} shows that when Assumption~\ref{AS:2} is satisfied and the proposed weighting scheme is used, the performance of self-supervised learning nearly matches that of fully-supervised learning. Fig.~\ref{fig:simulated_result} also highlights that using weighted self-supervised loss improves the imaging quality even when Assumption~\ref{AS:2} is not satisfied (i.e., the union of all the sampling masks doesn't cover the full k-space). For instance, note how the brain tissue highlighted using a green arrow is blurry for \emph{Full-rank (unweighted)}, while \emph{Full-rank} can reconstruct it with fine details. Also note that while the settings $\mathcal{J}_\mathsf{full}$ provide better reconstruction performances compared to those of $\mathcal{J}_\mathsf{def}$ under the same losses, \emph{Rank-deficient} outperforms \emph{Full-rank (unweighted)}, highlighting the effectiveness of the weighted matrix $\Wbm$ for self-supervised learning. Table~\ref{tb:simulated_result} summarizes PSNR/SSIM values of ablation methods on the testing dataset, thus quantitatively corroborating the visual results.

Fig.~\ref{fig:simulated_plot} plots PSNR and SSIM values against training epochs on the validation set with $R = 8$. Fig.~\ref{fig:simulated_plot} shows that \emph{Full-rank} with weighted matrix has approximately the same PSNR/SSIM curve as the \emph{supervised} baseline. Fig.~\ref{fig:simulated_plot} also shows that using a weighting matrix $\Wbm$ in the self-supervised loss can significantly improve imaging quality, even in the setting $\mathcal{J}_\mathsf{def}$ where Assumption~\ref{AS:2} does not hold.

\subsection{Experimentally Collected In-Vivo Data}\label{sec:exp_real} 

\subsubsection{Dataset} Data acquisition was performed on a Siemens 3T Prisma scanner (Siemens Healthcare, Erlangen, Germany) with 64-channel Head/Neck coils. We collected images using the Sagittal T1 magnetization-prepared rapid gradient-echo (MPRAGE) sequence. The acquisition parameters were as follows: repetition time (TR) $=$ 2400 ms, echo time (TE) $=$ 2.62 ms, inversion time (TI) $=$ 1000 ms, flip angle (FA) $=$ 8 degrees, FOV $=$ 256$\times$256 mm, voxel size $=$ 1$\times$1$\times$1 mm, slices per slab = 176, slice and phase resolution $=$ 100\% and slice and phase partial Fourier off. A $2\times$ oversampling was used in the frequency encoding direction, and the asymmetric echo was turned on to allow short TE.
The sampling pattern is equispaced 1D Cartesian with ACS lines.
Upon the approval of our Institutional Review Board, we used brain MRI data from 14, 1, and 5 participants in this study for training, validation, and testing, respectively.
We acquired the training data with GRAPPA $=$ 2 in phase encoding (PE) direction with 24 ACS lines, the total acquisition time was 5 minutes and 35 seconds, and the raw measurements correspond to approximately 65\% sampling rate.
The validation and testing data were fully-sampled measurements acquired with GRAPPA turned off, and the total acquisition time was 10 minutes and 16 seconds.
We considered groundtruth as the \emph{root-sum-square (RSS)} reconstruction from the fully-sampled data.

Experiments used the acceleration factors of $R = 4$, $R = 6$, and $R = 8$, corresponding to the retrospectively sampling rate of $30\%$, $20\%$, and $16\%$, respectively. We obtained the two training measurements of the same subjects (\emph{i.e.,} $\ybm_i'$ and $\ybm_i$ in \eqref{equ:proposed_imging}) by allocating acquired Cartesian lines into two bins.
% , and the resulting measurements have the same Cartesian sampling pattern with the raw data. 
Note that no groundtruth data was used during training. We applied 1D Fourier transform on the $k_z$ dimension of the raw data and then reconstructed the images slice by slice. 
The raw measurement of each slice is of size $512 \times  256\times 64$ with $512 \times 256$ being $k_x \times k_y$ dimension and $64$ being the numbers of receiver coils. Note that the high number of receiver coils makes DU methods impractical due to the increase in the computation and memory complexity.

\begin{table}[!ht]
  \centering
  \small
  \begin{threeparttable}
  \renewcommand\arraystretch{1.1}
  \setlength{\tabcolsep}{2.1pt}
  \begin{tabular}{cccccccc}
  \toprule
  {Metrics}                   & \multicolumn{3}{c}{PSNR (dB)}       & \multicolumn{3}{c}{SSIM}  & \multirow{2}{*}{Mem\tnote{1}}  \\
  {Acceleration rate}             & $\times8$   & $\times6$   & $\times4$    & $\times8$   & $\times6$   & $\times4$ &   \\
  \cmidrule(rl){2-4}\cmidrule(rl){5-7}
  \emph{Zero-Filled}                & 16.86 &	17.61 &	19.30 & 0.698 &	0.735 &	0.802 & N/A \\
  \rowcolor{gray!20} \emph{TV}  & 26.57 &	30.53 & 38.54 &	0.862 &	0.913 &	0.971 & 1745 \\
  \emph{A2A} & 29.80 & 31.84 & 35.03 & 0.874 & 0.903 & 0.940 & 7859\\
  \rowcolor{gray!20} \emph{SSDU} & 31.10 & 32.21 & 36.43 & 0.895 & 0.920 & 0.961 & 21981 \\
  \emph{SelfDU} &  32.19 & 34.44 & 37.65 & 0.906 & 0.931 & 0.961 & 21981 \\
  \rowcolor{gray!20} \emph{SelfDEQ}$^\mathsf{w/o}$  & 33.80 & 36.05 & 39.22 & 0.928 & 0.949 & 0.973 & 9325 \\
  \emph{SelfDEQ} & {\bf 34.05} & {\bf 36.79} & {\bf 39.64} & {\bf 0.929} & {\bf 0.950} & {\bf 0.973} & 9325  \\
  \bottomrule 
  \end{tabular}
  \begin{tablenotes}
    \item[1] GPU memory demand for training (MB). \\
    \emph{SelfDEQ}$^\mathsf{w/o}$: SelfDEQ~\emph{(unweighted)}.
  \end{tablenotes}
  \end{threeparttable}
  \caption{Summary of the PSNR and SSIM values on the experimentally-collected data. This table highlights that SelfDEQ~can outperform several self-supervised methods, including those based on DU.}
  \label{tb:real_result}
\end{table}

\begin{figure}[!h]
  \centering 
  \includegraphics[width=.48\textwidth]{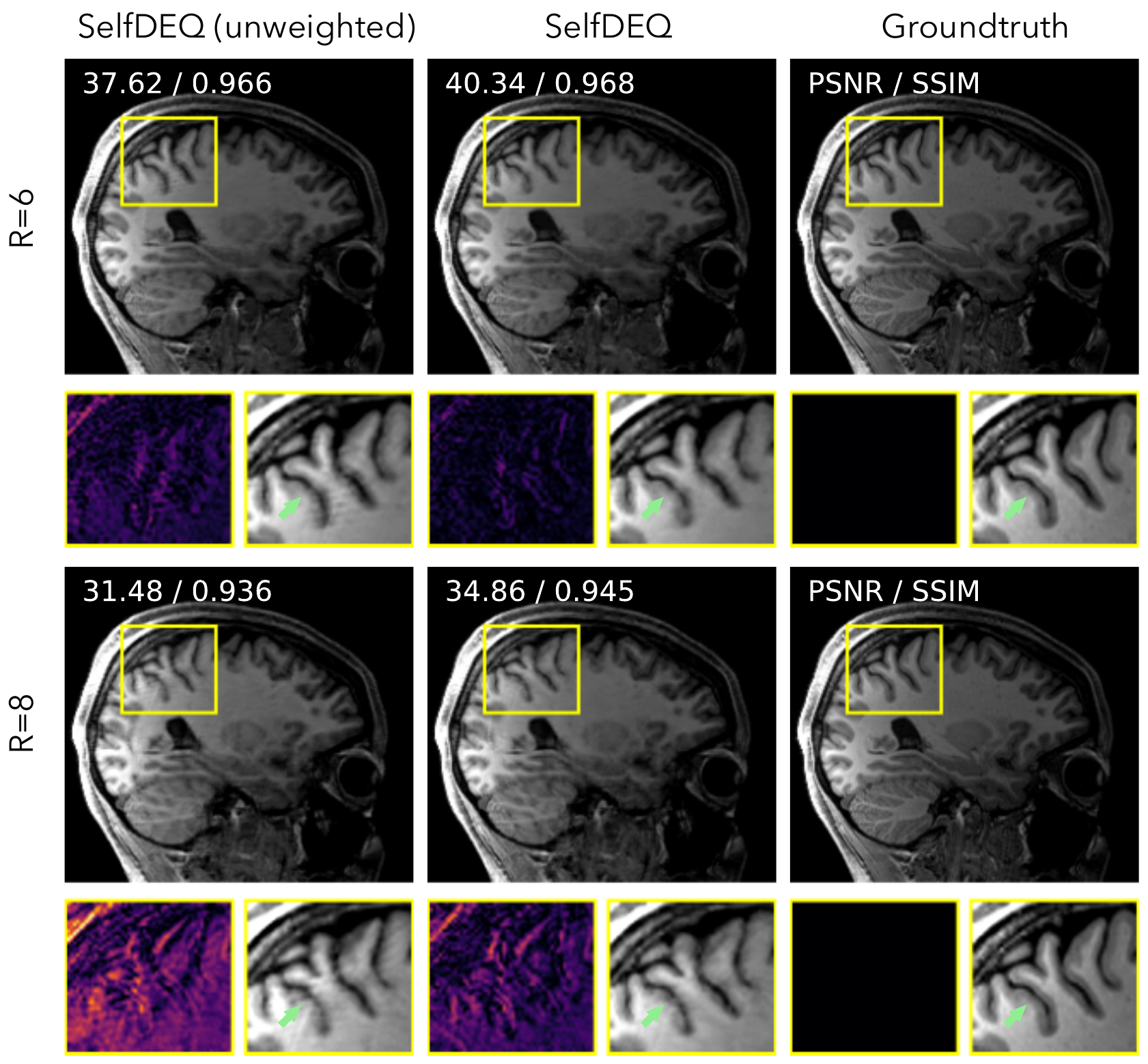}
  \caption{The reconstruction performance of SelfDEQ and its ablated variant that does not use the weighting matrix in the self-supervised loss. This figure shows results on acceleration factors $R =6$ and $R =8$, using the same image slice from the \emph{experimentally-collected} data. Note the improvement due to the  proposed weighting matrix $\Wbm$.}
  \label{fig:real_result_aba}
\end{figure} 

\subsubsection{Comparison} We compared SelfDEQ against several standard baseline methods and state-of-the-art self-supervised MRI methods. (a) \emph{Total Variation (TV)}: an optimization-based method using total variation regularizer in \eqref{equ:tv}. We optimized the trade-off parameter $\tau$ using grid search. (b) \emph{Artifact2Artifact (A2A)}~\cite{Liu.etal2020}: trains U-Net by mapping corrupted MR images of the same subject to each other. (c) \emph{SSDU}\footnote{SSDU implementation is available on GitHub: \url{https://github.com/byaman14/SSDU.}}~\cite{Yaman.etal2020}: a recent self-supervised method that trains a DU network by dividing each k-space MRI acquisition into two subsets and mapping them to each other. (d) \emph{SelfDU}: a DU network trained on the same DL architecture and using the \emph{weighted} loss function of SelfDEQ. We set DU iterations of \emph{SSDU} and \emph{SelfDU} to $7$, which is the maximum number achivable under memory constraints of our machine.
We also implemented SelfDEQ~\emph{(unweighted)} as an ablated method that trains SelfDEQ~on the self-supervised loss \emph{without} $\Wbm$.

% \begin{enumerate}[\ (a)]
%   \item \emph{Total Variation}: an optimization-based method with total variation regularizer, which is formulated in \eqref{equ:tv}. We optimized the trade-off parameter $\tau$ through grid searching.
%   \item \emph{Artifact2Artifact (A2A)}~\cite{Liu.etal2020}: trains a U-Net by mapping corrupted MR images of the same subject to each other.
%   \item \emph{SSDU}\footnote{SSDU implementation is available on GitHub: \url{https://github.com/byaman14/SSDU.}}~\cite{Yaman.etal2020}: a recent self-supervised method that trains a DU network by dividing each k-space MRI acquisition into two subsets and mapping them to each other.
%   \item \emph{SS-DUPGM}: a DU network trained on the same DL architecture and \emph{weighted} loss function for fair comparison.
% \end{enumerate}

\subsubsection{Discussion}
Fig.~\ref{fig:real_result} illustrates the reconstruction results of all
baseline methods on $R = 6$.
\emph{Total variation} suffers from detail loss due to the well-known ``staircase effect.'' While \emph{Artifact2Artifact} has better performance than \emph{Total variation} by learning the prior from data,
\emph{SSDU} and \emph{SelfDU} outperform it due to their model-based DU architectures. Overall, SelfDEQ achieves the best performance in artifact removal and sharpness. For instance, the reconstructed images obtained using SelfDEQ are sharper and have more fine details, especially in the brain tissues highlighted by green arrows. On the other hand, other methods show ghosting artifacts in their reconstructed images (see also zoom-in regions in Fig.~\ref{fig:real_result}).
Table~\ref{tb:real_result} summarizes the average PSNR and SSIM values of all the baseline methods on the testing dataset. The quantitative evaluations in Table~\ref{tb:real_result} show the superior performance achieved by SelfDEQ.
Table~\ref{tb:real_result} also provides the GPU memory requirements of each method for training, highlighting that SelfDEQ~can achieve better results with lower GPU memory demand than the DU-based methods.

Fig.~\ref{fig:real_result_aba} illustrates reconstruction results of SelfDEQ~and SelfDEQ\emph{ (unweighted)} on different acceleration factors and on the same image slice. Fig.~\ref{fig:real_result_aba} shows that using the weighted matrix $\Wbm$ in the self-supervised loss function can improve the imaging quality at different sampling rates.

\section{Discussion and Conclusion}

% \subsection{Generalization}
% Although this study focuses on CS-MRI, the proposed method can be naturally generalized to imaging problems where the forward operators consist of orthogonal transformation and subsampling matrix, such as the image inpainting problems.

\subsection{Applicability}
Practically obtaining training data for SelfDEQ is straightforward.
According to Assumption~\ref{AS:2}, it is sufficient to have a set of forward operators, where each operator subsamples the measurement domain, but their union over the training data covers the full space. For example, in MRI, one can implement a set of sampling masks illustrated in Fig.~\ref{fig:simulated_mask}, then randomly pick one of the sampling masks from the set during scanning.
In this example, individual undersampled measurements are still compatible with widely-used imaging techniques, such as GRAPPA~\cite{Griswold.etal2002} or ESPIRiT~\cite{Uecker.etal2014}.

\subsection{Future work} 
One benefit of SelfDEQ is its memory efficiency, which is well suited for large-scale imaging problems with high dimensional data.
In our experiments on real MRI data, we have applied SelfDEQ~on parallel MRI where the dimension of the receiver coils are high for conventional DU methods.
In future work, we will apply SelfDEQ on other high-dimensional data such as 4D free-breathing MRI~\cite{Eldeniz.etal2021} or 2D+time cardiac MRI~\cite{Pramanik.Jacob2022a}, where it is also challenging to obtain high-quality groundtruth training data.

\subsection{Conclusion}
This work presents SelfDEQ as a novel self-supervised learning framework for training model-based deep implicit neural networks for image reconstruction in accelerated MRI. The motivation behind SelfDEQ is to enable \emph{efficient} and \emph{effective} training of \emph{implicit networks} directly on undersampled and noisy MRI measurements without any groundtruth. The SelfDEQ framework consists of several synergistic elements: (a) a model-based implicit network that integrates measurement operators and CNN priors; (b) a self-supervised loss that accounts for sampling imbalances; (c) a Jacobian-free backward pass that leads to efficient training. We theoretical analysed SelfDEQ showing that it can do as well as the supervised learning. We tested the SelfDEQ framework on real MRI data, showing that it (i) outperforms recent DU based self-supervised methods; (ii) matches the performance of fully-supervised DEQ; and (iii) enables highly-accelerated data-collection in parallel MRI.

\section*{Acknowledgements}
Research reported in this publication was supported by the NSF CAREER award under CCF-2043134.

\newpage

\section*{Supplementary Materials}

Our main manuscript presents \emph{self-supervised deep equilibrium model (SelfDEQ)} as the first self-supervised MRI reconstruction framework for training model-based implicit neural networks from undersampled and noisy measurements. This supplementary document presents the details of our theoretical analysis. 

We use the same notations as in the main manuscript. We consider reconstruction of an image $\xbm\in\C^n$ from its noisy and undersampled measurement
\begin{equation}
\label{equ:imging_formulation}
    \ybm = \Mbm\Abm\xbm + \ebm,
\end{equation}
where $\ebm\in\C^m$ is a \emph{additive white Gaussian noise (AWGN)} vector, $\Abm\in\C^{n\times n}$ is a measurement matrix, and $\Mbm\in\{0, 1\}^{m\times n}$ is a diagonal sampling matrix.
We define the \emph{average weighting matrix} $\overline{\Wbm}$ to account for unbalanced sampling across multiple acquisitions in the training data
\begin{equation}
  \overline{\Wbm} = \mathsf{diag}(\overline{w_0}, \overline{w_1}, ..., \overline{w_{n}})\in\R^{n\times n}\ ,
\end{equation}
where $$
\overline{w_k} = \begin{cases}
  \frac{1}{\sqrt{{\E[{\Mbm'}^\Tsf\Mbm']}_{k,k}}}&{\sqrt{\E[{\Mbm'}^\Tsf\Mbm']_{k,k}}} \neq 0 \\
  0&\sqrt{{\E[{\Mbm'}^\Tsf\Mbm']}_{k,k}} = 0
\end{cases}\ .
$$
We define the \emph{weighting matrix} $\Wbm$ as a subsampled variant of $\overline{\Wbm}$ given an individual subsampling operator $\Mbm'$, which is used in our self-supervised loss function
\begin{equation}
    \Wbm = \Mbm'\overline{\Wbm}{(\Mbm'\overline{\Wbm})}^\Tsf\in\R^{m\times m}\ .
\end{equation}
Consider the training set of measurement pairs $\{\ybm_i, \ybm_i'\}_{i = 1}^N$ with each pair $\ybm_i, \ybm_i'$ corresponding to the same object $\xbm_i$
\begin{equation}
  \ybm_i = \Mbm_i\Abm\xbm_i + \ebm_i\ \text{and}\ \ybm_i'=\Mbm_i'\Abm\xbm_i+\ebm_i'\ .
\end{equation}
The value $N \geq 1$ denotes the number training pairs. 

The traditional DEQ gradient for the loss $\ell$ is given by
\begin{equation}
\label{equ:generic-DEQ}
    \begin{aligned}
       & \nabla\ell(\thetabm) = \mathsf{Real}\Big(\big(\nabla_\thetabm\Tsf_\thetabm(\xbmbar)\big)^\Hsf\bbm\Big)\\
       & \quad\text{where } \bbm = {\big(\Ibm-\nabla_\xbm\Tsf_\thetabm(\xbmbar)\big)}^{-\Tsf}{\Big[\frac{\partial\ell}{\partial\xbmbar}\Big]}^\Tsf\ .
    \end{aligned}
\end{equation}
Jacobian-free backpropagation (JFB) approximates \eqref{equ:generic-DEQ} as 
\begin{equation}
\label{equ:jfb-DEQ}
    \mathsf{JFB}_\ell(\thetabm) = \mathsf{Real}\Big(\big(\nabla_\thetabm\Tsf_\thetabm(\xbmbar)\big)^\Hsf{\Big[\frac{\partial\ell}{\partial\xbmbar}\Big]}^\Tsf\Big)\ .
\end{equation}
The analysis below shows that the JFB update using the self-supervised loss matches that using the supervised loss.

\begin{proposition}
  \label{PRO:new}
  When Assumption~2 is satisfited,
  $$
  \E\big[{(\Mbm'\Abm)}^\Hsf\Wbm\Mbm'\Abm\big]=\Ibm\ ,
  $$
  where the expectation is with respect to $p_{\Mbm}$.
\end{proposition}

\noindent
\emph{Proof:} 
Since Assumption 2 implies that $\E[\Mbm']_{k,k} \neq 0$
\begin{equation}
    \overline{w_k} = \frac{1}{\sqrt{{\E[{\Mbm'}^\Tsf\Mbm']}_{k,k}}}\ .
\end{equation}
Since ${\Mbm'}^\Tsf\Mbm'\in\{0,1\}^{n\times n}$ and $\overline{\Wbm}\in\R^{n\times n}$ are both diagonal matrices, we have
\begin{equation}
\label{equ:wm_i}
    \begin{aligned}
       & \E[{\Mbm'}^{\Tsf}\Wbm\Mbm'] \\ & = \E[{\Mbm'}^{\Tsf}\Mbm'\overline{\Wbm}\ \overline{\Wbm}^\Tsf{\Mbm'}^\Tsf\Mbm'] \\ 
       & = \overline{\Wbm}\ \overline{\Wbm}^\Tsf\E[{\Mbm'}^{\Tsf}\Mbm'{\Mbm'}^\Tsf\Mbm'] \\ 
      %& = \mathsf{diag}(\overline{w}_k^2 * \E[{\Mbm'}^\Tsf\Mbm']_{k,k})\ \forall k=1...n \\
      & = \Ibm\ . 
    \end{aligned}
\end{equation}
Now we can establish the desired result
\begin{equation}
  \begin{aligned}
    & \E\big[{(\Mbm'\Abm)}^\Hsf\Wbm\Mbm'\Abm\big]  = \Abm^\Hsf\E\big[\Mbm'^\Tsf\Wbm\Mbm'\big]\Abm \\
    & = \Abm^\Hsf\Abm = \Ibm\ .
  \end{aligned}
\end{equation} 
where the second equation are due to \eqref{equ:wm_i}, and the last equation is because $\Abm$ is an orthogonal matrix. 

\begin{theorem}
Under Assumptions~1-2, the JFB update of the weighted self-supervised loss ($\ell_\mathsf{self}$) is equivalent to its supervised counterpart ($\ell_\mathsf{sup}$), namely we have that
\begin{equation}
    \begin{aligned}
        & \mathsf{JFB}_{\ell_\mathsf{self}}(\thetabm) = \mathsf{JFB}_{\ell_\mathsf{sup}}(\thetabm)\ .
    \end{aligned}
\end{equation}
where 
\begin{equation}
    \ell_\mathsf{sup}=\E\left[\frac{1}{2}\norm{\xbmbar - \xbm}_2^2\right]
\end{equation}
and
\begin{equation}
    \ell_\mathsf{self}=\E\left[\frac{1}{2}\norm{\Mbm'\Abm\xbmbar - \ybm'}_{\Wbm}^2\right]\ .
\end{equation}
The vector $\xbmbar = \mathsf{T}_\thetabm(\xbmbar, \ybm)$ is the fixed-point of $\mathsf{T}_\thetabm$ for $\ybm$.
\end{theorem}

\noindent
\emph{Proof:} 
In order to simplify the notations in the following analysis, we directly use complex valued quantities and assume that the real part is taken at the end.

The supervised update $\mathsf{JFB}_{\ell_\mathsf{sup}}(\thetabm)$ is given by:
\begin{equation}
\label{equ:jfb_sup}
    \begin{aligned}
        & \mathsf{JFB}_{\ell_\mathsf{sup}}(\thetabm) = \E\Big[\big(\nabla_\thetabm\Tsf_\thetabm(\xbmbar)\big)^\Hsf{\Big[\frac{\partial\ell_\mathsf{sup}}{\xbmbar}\Big]}^\Tsf\Big] \\
%        =\ & \E\Big[\big(\nabla_\thetabm\Tsf_\thetabm(\xbmbar)\big)^\Tsf{\Big[\frac{\partial\ \frac{1}{2}\norm{\xbmbar - \xbm}_2^2}{\partial\xbmbar}\Big]}^\Tsf\Big] \\
        & =  \E\Big[\big(\nabla_\thetabm\Tsf_\thetabm(\xbmbar)\big)^\Hsf(\xbmbar - \xbm)\Big]\ .
    \end{aligned}
\end{equation}
On the other hand, we can re-write the weighted self-supervised update as $\mathsf{JFB}_{\ell_\mathsf{self}}(\thetabm)$:
    \begin{align}
        \label{equ:self_jfb}& \mathsf{JFB}_{\ell_\mathsf{self}}(\thetabm) = \E\Big[\big(\nabla_\thetabm\Tsf_\thetabm(\xbmbar)\big)^\Hsf{\Big[\frac{\partial\ell_\mathsf{self}}{\partial\xbmbar}\Big]}^\Tsf\Big] \\
        \nonumber& = \E\Big[\big(\nabla_\thetabm\Tsf_\thetabm(\xbmbar)\big)^\Hsf\big(\Hbm'^\Hsf(\Hbm'\xbmbar-\sqrt{\Wbm}\ybm')\big)\Big] \\
        \nonumber& = \E\Big[\big(\nabla_\thetabm\Tsf_\thetabm(\xbmbar)\big)^\Hsf\E\big[\Hbm'^\Hsf(\Hbm'\xbmbar-\sqrt{\Wbm}\ybm')\big|\xbm,\Mbm,\ebm\big]\Big]\ ,
\end{align}
where $\Hbm'{\defn} \sqrt{\Wbm}\Mbm'\Abm$. The last equation is true since $\nabla_\thetabm\Tsf_\thetabm(\xbmbar)$ is deterministic when conditioned on $\xbm$, $\Mbm$ and $\ebm$. We also have that 
\begin{align}
        \label{equ:self_01} & \E\big[\Hbm'^\Hsf(\Hbm'\xbmbar-\sqrt{\Wbm}\ybm'\big)\big|\xbm,\Mbm,\ebm\big] \\
        \nonumber&= \E\big[\Hbm'^\Hsf\big(\Hbm'\xbmbar-\sqrt{\Wbm}( \Mbm'\Abm\xbm+\ebm')\big)\big|\xbm,\Mbm,\ebm\big] \\
        \nonumber& = \E\big[\Hbm'^\Hsf\big(\Hbm'(\xbmbar-\xbm)+\sqrt{\Wbm}\ebm'\big)\big|\xbm,\Mbm,\ebm\big] \\
        \nonumber&= \E\big[\Hbm'^\Hsf\Hbm'(\xbmbar-\xbm)\big|\xbm,\Mbm,\ebm\big]+\E\big[\Hbm'^\Hsf\sqrt{\Wbm}\ebm'\big|\xbm,\Mbm,\ebm\big],
\end{align}
where in the second row we used $\ybm = \Mbm'\Abm\xbm + \ebm'$. The first term in \eqref{equ:self_01} can also be expressed as
\begin{equation}
\label{equ:self_0}
    \begin{aligned}
        & \E\big[\Hbm'^\Hsf\Hbm'(\xbmbar-\xbm)\big|\xbm,\Mbm,\ebm\big] \\
%        &= \E\big[\Hbm'^\Hsf\Hbm'\big|\xbm,\Mbm,\ebm\big](\xbmbar-\xbm) \\
        &= \E\big[\Hbm'^\Hsf\Hbm'\big](\xbmbar-\xbm) \\
        &= \xbmbar-\xbm\ ,
    \end{aligned}
\end{equation}
where the second equation is due to independence of $\Mbm'$ from $\xbm,\Mbm$ and $\ebm$, and the last equation is due to Proposition 1. The second term of \eqref{equ:self_01} can be expressed as
\begin{equation}
\label{equ:self_1}
    \begin{aligned}
       & \E\big[\Hbm'^\Hsf\sqrt{\Wbm}\ebm'\big|\xbm,\Mbm,\ebm\big] \\
       & = \E\big[\Hbm'^\Hsf\sqrt{\Wbm}\ebm'\big] = \E\big[\Hbm'^\Hsf\sqrt{\Wbm}\big]\E\big[\ebm'\big]  = \bm{0}\ ,
    \end{aligned}
\end{equation}
where the first equation is due to the independence of $\Mbm'$ in $\Hbm'$ and $\ebm'$ from $\xbm,\Mbm$ and $\ebm$, the second equation is due to the independence of $\Mbm'$ from $\ebm'$, and the last equation is due to $\ebm'\sim\Ncal(0,\sigma^2\Ibm)$.
Combining \eqref{equ:self_jfb}, \eqref{equ:self_01}, \eqref{equ:self_0} and \eqref{equ:self_1}, we get 
\begin{equation}
\label{equ:jfb_self_1}
    \mathsf{JFB}_{\ell_\mathsf{self}}(\thetabm) = \E\Big[\big(\nabla_\thetabm\Tsf_\thetabm(\xbmbar)\big)^\Hsf(\xbmbar - \xbm)\Big]\ ,
\end{equation}
which establishes the desired results.

\end{document}